\documentstyle[12pt]{article}
\begin{document}
\begin{titlepage}
\pagestyle{empty}
\baselineskip=21pt
\rightline{hep-ph/9506362}
\rightline{Astropart.Phys.(to appear)}
\vskip .2in
\begin{center}
{\large{\bf Astrophysical and cosmological considerations \\
 on a string dilaton of a least coupling \\}} 
\end{center}
\vskip .1in
\begin{center}
C. E. Vayonakis

{\it Physics Department, University of Ioannina,}

{\it GR-451 10 Ioannina, Greece} \footnote {Permanent address}

{\it and} 

{\it Physics and Astronomy Subject Group,}

{\it University of Sussex, Brighton BN1 9QH, U.K.}

\vskip .1in

\end{center}
\vskip .5in
\centerline{ {\bf Abstract} }
\baselineskip=18pt
String-loop effects may generate very weak matter couplings for a
(massless) dilaton. We examine limits on the shift of such a dilaton
toward its present equilibrium value from big-bang nucleosynthesis and the binary
pulsar. On the other hand, the approach of the dilaton toward its present value
can be realized very early in the universe in a quick and efficient way if an
inflationary period is present. We comment briefly on some implications.

\noindent

\end{titlepage}
\baselineskip=18pt

\def\la{{{\lower 5pt\hbox{$<$}} \atop {\raise 5pt\nbox{$\sim$}}}~}
\def\ga{{{\lower 2pt\hbox{$>$}} \atop {\raise 1pt\hbox{$\sim$}}}~}
\def\mtw#1{m_{\tilde #1}}
\def\tw#1{${\tilde #1}$}
\def\beq{\begin{equation}}
\def\eeq{\end{equation}}
\def\lessim{\lower0.6ex\hbox{$\,$\vbox{\offinterlineskip
\hbox{$<$}\vskip1pt\hbox{$\sim$}}$\,$}}
\def\grtsim{\lower0.6ex\hbox{$\,$\vbox{\offinterlineskip
\hbox{$>$}\vskip1pt\hbox{$\sim$}}$\,$}}

\noindent{\bf I.\quad Introduction}
\medskip
\nobreak

String theory provides, at present, the only scheme capable to give a
promising unified picture of gauge particle interactions and quantum
gravity. Within this theory, there is the usual tensor Einstein graviton
always intimately combined with a universal partner: the dilaton. The
mass of the dilaton is presently unkown. Its couplings to matter a
priori generate deviations from Einstein gravity. In fact, short
distance tests of Newtonian gravity can be used to put bounds on the
dilaton mass. Based on $E\ddot{o}tv\ddot{o}s$-type experiments and satellite
observations one can place a bound $m_{\phi} \grtsim 10^{-4}\rm eV$
\cite{ekow}. On the other hand, Cavendish-type experiments, which probe
deviations from the inverse-square law, can be used to put a lower bound
on the mass of a tree-level string dilaton $m_{\phi} \grtsim 10^{-3}
\rm eV$ \cite{dp1}. Present conventional wisdom is to have a non-vanishing
dilaton mass \cite{gvg}, possibly connected with a (dynamical) supersymmetry
breaking mechanism \cite{dccqr}.

A very interesting proposal, however, has been put forward in ref.\cite{dp1}
(see also \cite{dp2}), offering the alternative possibility of a massless
dilaton. In this, it is pointed out that non-perturbative string-loop
effects (associated with worldsheets of arbitrary genus in indermediate
string states) may naturally reconcile the existence of a massless
dilaton with experimental data, if the various couplings of the
dilaton field to the other fields exhibit a certain kind of universality
(as the tree-level dilaton couplings do). Then, under the assumption
that the different coupling functions $B_i(\Phi)$ have extrema at some
point $\Phi = \Phi_m$, the study of the cosmological evolution of a
general graviton-dilaton-matter system during the radiation- and the
matter -dominated eras shows \cite{dp1} that the dilaton is cosmologically
attracted toward the value $\Phi_m$, where it decouples from matter.
This illustrates a ``least coupling'' mechanism in the sense that the
universe is attracted  to dilaton values extremizing the strengths of its
interactions. The attaction mechanisms discussed in \cite{dp1} (see also
\cite{dn}), driving the vacuum expectation value of the dilaton toward values at
which it decouples from matter, are not however perfect
attractors. It remains, then, the important question of giving quantitative
independent estimates of these residual coupling strengths of the
dilaton.
\bigskip

\noindent{\bf II.\quad The relaxation mechanism paradigm}
\medskip
\nobreak

To be concrete, let us summarize in this section the proposal of ref.\cite{dp1}
by considering the effective action for the string massless modes
$$
S = {\int}d^4 x \sqrt{\hat{g}} B(\Phi) \left\{ \frac{1}{\alpha'}\hat{R}
+ \frac{1}{\alpha'}\left[4\hat{\nabla^{2}}\Phi - 4(\hat{\nabla}\Phi)^2\right] - 
\frac{k}{4}\hat{F}^{2} - \bar{\hat{\psi}}\hat{\nabla}\hat{\psi} - 
\frac{1}{2} (\hat{\nabla}\hat{\sigma})^2 - \hat{V}(\hat{\sigma})
\right\}  \eqno{(1)}$$
where the dilaton coupling function $B(\Phi)$ appearing as a common
factor in front admits a series expansion of the form
$$
B(\Phi) = e^{-2\Phi} + c_0 + c_1 e^{2\Phi} + c_2 e^{4\Phi} + ... . 
\eqno{(2)}$$
The first term in the expansion (2) is the string tree-level
contribution (corresponding to spherical topology for indermediate
worldsheets), which is known to couple the dilaton in a universal
multiplicative manner to all other fields and is derived from the fact
that $g_s = exp(\Phi)$ plays the role of the string coupling constant
\cite{ft}. The remaining terms represent the string-loop effects: the genus-n
string-loop contributions to any string transition amplitude contains a
factor $g_s^{2(n-1)} = exp[2(n-1)\Phi]$. At the present stage of
development of string theory, little is known about the global behaviour
of dilaton coupling function $B(\Phi)$, apart from the fact that in the
limit $\Phi \rightarrow - \infty$ ($g_s \rightarrow 0$) it should admit
a series expansion in powers of $g_s^2 = exp[2\Phi]$. In general, one
expects to have various dilaton coupling functions $B_i(\Phi)$ as
coefficients in front of each term appearing in the action. For the
cosmological attraction mechanism discussed in \cite{dp1} to work, it suffices
that a universality condition must be fulfilled, namely that the various
coupling functions $B_i(\Phi)$ must all admit a local maximum at some
common value $\Phi_m$. For the action (1) this universality is
guaranteed by the factorization of the common function $B(\Phi)$. In
that case, string-loop effects as in (2) can allow this function to
admit a local maximum. In the following we assume such a common function
$B(\Phi)$.
It might be stressed that the desired universality could be attributed
to the conjectured S-duality of the string theory \cite{filq}, namely the
minimum/maximum (weak/strong) coupling constant duality $g_s \rightarrow
{1 \over{g_s}}$ corresponding to the discrete symmetry $\Phi \rightarrow
- \Phi$. This property is intrinsically non-perturbative and would
ensure that all physical quantities have extremum at $\Phi_m = 0$. A
similar duality symmetry is known to hold for the other gauge-neutral
(massless) scalar fields present in string theory, namely those
associated with the compactified dimensions and known as moduli. The
latter symmetry, called T-duality or target-space duality \cite{gpr}( a
minimum/maximum length duality $R \rightarrow {\alpha' \over{R}}$) is manifest
order by order in string perturbation theory (though non-perturbative on the
string worldsheet). It is true that, unlike T-duality, S-duality has
not yet been proved to be a string theory symmetry, although some
interesting and non-trivial tests can be carried out \cite{sen}. There
is, in any case,
 some evidence coming from different directions that S- and
T-duality are part of a larger unified symmetry structure \cite{bakas}. Under
the spirit of these considerations, it was conjectured \cite{dp1,dp2} that the
proposed attractor mechanism could be invariably applied to the dilaton as well
as the various moduli fields.
      
It is convenient to transform the action (1) by introducing several
$\Phi$-dependence rescalings. In particular, one can put the gravity
sector (graviton, dilaton) and the matter one (fermions, gauge fields,
scalars) into a standard form by replacing the original "string frame"
metric ${\hat g}_{\mu\nu}$ by a conformally related "Einstein
frame" metric
$$
g_{\mu\nu} = CB (\Phi){\hat g}_{\mu\nu}
\eqno{(3)}$$
the original field $\Phi$ by a canonical scalar field
$$
\phi = \int d\Phi \left[ {{3 \over{4}}\left( {B' \over{B}}\right)^2 +
2{{B'} \over{B}} + 2} \right] ^{1/2}
\eqno{(4)}$$
( where prime denotes d/d$\Phi$ ) and the original Dirac field $\hat
\psi$ and scalar field $\hat \sigma$ by
$$
\psi = C^{-3/4}B^{-1/4}{\hat \psi},   \sigma = C^{-1/2}{\hat \sigma}
\eqno{(5)}$$
The transformed action reads
$$
S = {S_{gr}} \left(g,\phi \right) + {S_m} \left(\psi,A,\sigma,... \right)
\eqno{(6.a)}$$
$$
S_{gr} = \int d^4 x \sqrt{g} \left[ {1 \over{4q}}R - {1\over{2q}}(\nabla
\phi)^2 \right]
\eqno{(6.b)}$$
$$
S_m = \int d^4 x \sqrt{g} \left[ - {\bar \psi}\nabla \psi - {k
\over{4}}B(\phi)F^2 - {1\over{2}}(\nabla \sigma)^2 -
{B^{-1}}(\phi)V(\sigma) + ... \right]
\eqno{(6.c)}$$
where $q = 4\pi {\bar G} = \frac{1}{4}C\alpha'$ with $\bar{G}$ denoting a bare
gravitational coupling constant and $B(\phi) \equiv B \left[ \Phi(\phi)
\right]$. The constant C can be chosen so that the string units and the
Einstein units coincide at the present cosmological epoch $CB(\Phi_0) =
1$.
The relation of string models to the observed low-energy world is
through the dilaton dependence (from (6.c)) of the gauge coupling
constants
$$
g^{-2} (\phi) = kB(\phi)
\eqno{(7)}$$
Furthermore, one still needs to take into account the quantum effects of
the light modes between the string scale in Einstein units
$$
{\Lambda_{s}} (\phi) = C^{-1/2}B^{-1/2}(\phi)\hat{\Lambda_{s}}
\eqno{(8)}$$
and some (low) observational scale. In eq.(8) ${ \hat \Lambda_s} \simeq
3 \times 10^{17}\rm GeV$ \cite{kaplunovsky} is the (string frame) string unification
scale $\sim (a')^{-1/2}$. So, for example, the renormalization group
implies that, at the one-loop level, the QCD mass scale $\Lambda_{QCD}$ is
exponentially related to the inverse of $g_3$ gauge coupling constant and
through the above dependense we have
$$
\Lambda_{QCD} (\phi) \sim \Lambda_{s} (\phi) exp \left( -8{\pi}^{2}{b_3}^{-1}{g_3}^{-2}
\right) = C^{-1/2}B^{-1/2}(\phi)exp \left[ 
-8{\pi}^{2}{b_3}^{-1}{k_3}B(\phi) \right] {\hat \Lambda_s}
\eqno{(9)}$$
Later on we will need the dilaton dependence of the ( Einstein frame)
QCD part of the mass of the nucleons, which is given by some pure number
times $\Lambda_{QCD} (\phi)$. Furthermore, the mass ( in Einstein units) of
any type of particle, labelled A, will depend on $\phi$ only through the
function $B(\phi)$
$$
m_A (\phi) = m_A \left[ B(\phi) \right]
\eqno{(10)}$$
We assume a form dependence as suggested by (9), that is
$$
m_A (\phi) = \mu_A B^{-1/2}(\phi)exp \left[ -8{\pi}^2 \nu_A B(\phi)
\right]{\hat \Lambda_s}
\eqno{(11)}$$
with $\mu_A $, $\nu_A$ pure numbers $\sim O(1)$. This is in fact the case in
models in which the electroweak gauge symmetry is broken radiatively,
more specifically in the no-scale supergravity models \cite{cfkn}.

Upon studying the cosmological evolution of the graviton-dilaton-matter
system,the authors of ref.\cite{dp1} find that the dilaton vacuum expectation
value $\phi$ is dynamically driven toward some finite value $\phi_m$,
where the mass functions $m_A (\phi)$ reach a local minimum
corresponding to a local maximum  of $B(\phi)$. The important quantities
in the proposal under discussion are the attraction shift $\Delta \phi
\equiv (\phi - {\phi_m})$ of the $\phi$ toward the value $\phi_m$ during the
evolutionary history of the universe and the coupling strength
$$
\alpha_A (\phi) = {\partial \over{\partial \phi}}{ln m_A(\phi)}
\eqno{(12)}$$
of the dilaton to the A-type particles. This is seen from the field equation for
$\phi$, which is 
$$
\nabla^2 \phi = - q {\Sigma}
\eqno{(13)}$$
where the source term is 
$$
\Sigma = g^{-1/2}{\partial \over{\partial \phi}}{S_m}
\eqno{(14)}$$
Upon writing for the material content of the universe the action
$$
S_m = - \sum_{A} \int m_A(\phi) d \tau_A
\eqno{(15)}$$
we have
$$
\Sigma = \sum_{A} \alpha_A (\phi) T_A
\eqno{(16)}$$
where $T_A = - {\rho}_A + 3p{_A}$ is the trace of the $A$-type contribution to
the total energy-momentum tensor $T^\mu\nu = \sum_{A} {T_A}^\mu\nu$. Thus,
$\alpha_A (\phi)$ is the coupling strength of the dilaton to the A-type
particles and its square ${\alpha_A} ^2 (\phi)$ will appear in all
quantities involving interactions mediated by exchanges of dilatons (
in the same way as $g^2$ appears in all gauge interactions).

The main parameter determining both the efficiency of the cosmological
relaxation of $\phi$ toward $\phi_m$ and the coupling strength
${\alpha}_A (\phi)$ is the curvature $\kappa$ of the function $ln B(\phi)$
near its maximum at $\phi_m$. In the quadratic approximation we have
$$
ln B(\phi) \simeq const. - {1 \over 2} {\kappa} (\phi - \phi_m)^2
\eqno{(17)}$$
Then, from (11) and (12) one gets
$$
\alpha_A (\phi) \simeq \beta_A (\phi - \phi_m)
\eqno{(18)}$$
with $\beta_A = \kappa ln {e^{1/2}{\hat \Lambda_s}/{m_A}} \simeq 40\kappa
(\kappa \sim O(1))$.

As discussed by the authors of ref.\cite{dp1}, because of the steep dependence
of $m_A (\phi)$ on $B(\phi)$ [see eq.(11)], during the
radiation-dominated era each time the cosmic temperature T becomes of
the order of the mass threshold $m_A$ of some particle, $\phi$ is
attracted towards $\phi_m$ by a factor $\sim 1/3$. Besides mass
thresholds, the electroweak and the QCD phase transitions provide other
possible attractor processes during the same period. During the
subsequent matter-dominated era, $\phi$ is further attracted toward
$\phi_m$ by a factor roughly proportional to $Z_0 ^{-3/4}$, where $Z_0
\simeq 1.3 \times 10^4$ is the redshift seperating us from the end of the
radiation era. All this leads to a present value $\phi_0$ near but
different from $\phi_m$ by a very small amount $\Delta\phi_0 = (\phi_0 - \phi_m)$
and to a cosmologically relaxed dilaton coupling to matter around us
with a strength 
$$
\alpha_A (\phi_0) = {\partial \over{\partial \phi}}{ln m_A
(\phi)}|_{\phi=\phi_0} \simeq \beta_A (\phi_0 - \phi_m )
\eqno{(19)}$$
>From the evolution of the dilaton during the two well-separated eras of
classical cosmology, the authors of ref.\cite{dp1} conclude that today $(\phi_0
- \phi_m ) \simeq 10^{-9}$. Such a very small amount implies, of course,
very small deviations from general relativity. In particular, all
deviations from Einstein's theory ( including post-Newtonian deviations
measured by the two Eddington parameters $1 - \gamma_{Edd}$ and $\beta_{Edd}
- 1$, residual cosmological variation of the coupling constants and
violation of the weak equivalence principle) are proportional to a small
factor $(\phi_0 - \phi_m )^2$ coming from the exchange of a $\phi$
particle. The previous estimate gives a small factor $(\phi_0 - \phi_m
)^2 \simeq 10^{-18}$. The present high-precision tests of the
equivalence principle reach the $10^{-12}$ level ( corresponding to
$(\phi_0 - \phi_m )_{obs} \simeq 10^{-6}$ in the context of the present
model).

In the next, we are looking for some independent constraints for the
relaxation shift $\Delta\phi = (\phi - \phi_m )$ toward its present
value $\Delta\phi_0 = (\phi_0 - \phi_m )$. We choose two
well-established areas of cosmology and astrophysics, namely big-bang
nucleosynthesis and binary pulsars. The first offers the possibility to
costrain the relaxation shift $(\phi - \phi_m )$ from the time of
nucleosynthesis till present, while the second can provide an
independent estimate of the present value of $\alpha_A (\phi_0) \simeq
\beta_A (\phi_0 - \phi_m)$. We will also examine the effect that an
inflationary epoch could induce on the relaxation shift under discussion.
\bigskip

\newpage
\noindent{\bf III. Nucleosynthesis constraint}
\medskip
\nobreak

Let us first come to nucleosynthesis. Particle physics models are mostly
constrained by the mass fraction of primordial $^4$He, usually denoted
by $Y_p$. The light element abundances including $^4$He ( i.e. D,
$^3$He, $^4$He, $^7$Li) are mainly sensitive to the baryon-to-photon
ratio $\eta = {n_N}/{n_\gamma}$ within the standard model \cite{wsso}.
Consistency between theory and observations restrict $\eta$ to be in the
range $3.1 \times {10^{-10}} \lessim \eta \lessim 3.9 \times {10^{-10}}$. For $^4$He
it is found that \cite{steigman}
$$
Y_p = 0.232 \pm 0.003 \pm 0.005
\eqno{(20)}$$
where the uncertainties are $1\sigma$ statistical and systematic
uncertainties, respectively.
The $^4$He abundancy is mainly determined by the neutron-to-proton ratio
just before nucleosynthesis
$$
{n \over p} \simeq e^{-\Delta m_N/T_f}
\eqno{(21)}$$
where $\Delta m_N = 1.29 {\rm MeV}$ is the neutron-proton mass
difference and $T_f \sim 1 {\rm MeV}$ is the freeze-out temerature of
the weak interactions. Actually, the ratio is slightly altered due to
free neutron decays between $T_f$ and the onset of nucleosynthesis at
about $T \sim 0.1 {\rm MeV}$. The temperature $T_f$ is determined by the
competition between the weak interaction rate and the expansion rate of
the universe 
$$
{G_F}^2 {T_f}^5 \simeq \sqrt{G N} {T_f}^2
\eqno{(22)}$$
where N is the total number of relativistic particle species and $G_F$,
$G \simeq {\bar G}$ are Fermi's weak interactions and Newton's gravitational
coupling constants, respectively. The $^4$He abundance is then estimated
to be 
$$
Y_p \simeq {2 (n/p) \over {1 + (n/p)}}
\eqno{(23)}$$
Therefore, changes of $Y_p$ due to changes in $T_f$ and $\Delta m_N$ are
approximately 
$$
{\Delta Y_p \over Y_p} \simeq \left ( {\Delta T_f \over T_f} - {\Delta
(\Delta m_N) \over \Delta m_N} \right)
\eqno{(24)}$$
These changes have been discussed in the past in various settings \cite{kpw}
(ours is closer to the last of ref.\cite{kpw}). In the theory under
consideration, the dependence of the various quantities of interest on
dilaton ( through equations (7)-(11) and (17)) will induce changes on
them, which will then constrain the shift $\Delta \phi_{BBN} \equiv (\phi -
\phi_m)_{BBN}$ of $\phi$ toward $\phi_m$ from the time of big-bang
nucleosynthesis to today.																			 
Changes in $T_f$ are derived from changes in $G_F$ and G ( assuming N
fixed, see eq.(22)). Since we are working in the ``Einstein-frame'', we do
not consider changes in the gravitational coupling G. Changes in the
Fermi coupling $G_F$ can be deduced by taking $G_F(\phi) \sim {m_A}^{-2}
(\phi)$. Since we are only interested in getting the shift of $\phi$
toward $\phi_m$ from the time of nucleosynthesis till present, we take
that initially $(\phi - \phi_m) \sim O(1)$. We then find 
$$
{\Delta T_f \over T_f} \simeq -50 \Delta \phi_{BBN}
\eqno{(25)}$$
Changes in the neutron-proton mass difference $\Delta m_N$, on the other
hand, are derived from changes in both the $\Lambda_{QCD}$-dependent
contribution and the fermion (quark) mass contribution to nucleon
masses, written as \cite{leutwyler}
$$
\Delta m_N \sim c_1 {\alpha_{em}(\phi)}{\Lambda_{QCD}(\phi)} +
c_2{\upsilon(\phi)}
\eqno{(26)}$$
where $c_1$, $c_2$ are dimensionless constants and $\upsilon(\phi) \sim
{G_F}^{-1/2}(\phi)$ is the weak scale vacuum expectation value.
Replacing the numerical values of $c_1$, $c_2$ \cite{leutwyler} and taking the dilaton
dependence of the quantities as prescribed within the scheme we
consider, we find under the previous assumptions that 
$$
{\Delta (\Delta m_N) \over \Delta m_N} \simeq 50 \Delta \phi_{BBN}
\eqno{(27)}$$
So, we have a net change in $Y_p$
$$
| {\Delta Y_p \over Y_p} | \simeq 100 \Delta \phi_{BBN}
\eqno{(28)}$$
>From the range (20) of values for $Y_p$ we see that consistency with
big-bang nucleosynthesis requires 
$$
\Delta \phi_{BBN} \lessim 10^{-5}
\eqno{(29)}$$
The derived bound is a kind of ``phenomenological'' constraint given the
successful frame of the big-bang nucleosynthesis and shows the expected
shift of the dilaton field from that time up to the present.
\bigskip

\newpage
\noindent{\bf IV.\quad  Binary pulsar constraint} 
\medskip
\nobreak

Pulsars, in general \cite{blandford}, and binary pulsars, in particular
\cite{pk}, constitute a unique physics laboratory and can, most probably,
provide the ultimate test of gravity theories \cite{will,damour}. The point
which is of interest to us here is that the gravitational radiation and the
induced change in the period of binary pulsars, most notably the PSR 1913 +
16  \cite{ht}, are so well described within general relativity that any
deviations from it are severely constrained.

Different metric theories of gravity predict different possible types (
monopole, higher multipole ) of gravitational radiation emitted by a
given source, here the binary pulsar. This can be studied by analysing
the effects of gravitational radiation reaction on the source, here the
energy loss and the induced change in the period of the binary pulsar.
Denote by $m_1$, $m_2$ the masses of the pulsar and the companion, $m =
m_1 + m_2$ the total mass, $\mu = m_1 m_2 / m$ the reduced mass, r the
orbital separation and v the relative velocity. Then, the rate of energy
loss of the system due to the combined effect of quadrupole and monopole
gravitational radiation can be written, following ref.\cite{will},
$$
{dE \over dt} = - < \frac{\mu^{2} m^2} {r^{4}} {8\over{15}} ( k_1 v^2 - k_2
\dot{r^{2}} ) >
\eqno{(30)}$$
This energy loss induces a decrease in the orbital period $P_b$  given
by Kepler's third law
$$
\dot P_b / P_b = - {3 \over2}{dE \over dt} / E
\eqno{(31)}$$
Carrying out the average over one orbit using Keplerian orbital formulae
gives the expressions \cite{will}
$$
{dE \over dt} = - {32 \over 5} ( {\mu \over m} )^2 ( {m \over \alpha} )^5 F(e)
\eqno{(32)}$$

$${\dot P_b \over P_b} = - {96 \over 5} ({ \mu m^2 \over {\alpha}^4} ) F(e)
\eqno{(33)}$$
where $\alpha$ and e are the orbital semi-major axis and eccentricity,
and
$$
F(e) = {1 \over 2} \left [ k_1 (1 + {7 \over 2}e^2 + {1 \over 2}e^4) -
k_2 ({1 \over 2}e^2 + {1 \over 8}e^4) \right ](1 - e^2)^{-7/2}
\eqno{(34)}$$

Now, within general relativity $k_1 = 12, k_2 = 11$ and formulae
(32)-(34) constitute the quadrupole formula for the emission of
gravitational energy \cite{pm}. In that case, for the parameters of the binary
pulsar PSR 1913 + 16 formula (33) predicts \cite{will} $\dot P_b ^{GR} = -
\left ( 2.40243 \pm 0.00005 \right ) \times 10^{-12}$. 
The observed value is $\dot P_b ^{OBS} = -
\left ( 2.408 \pm 0.010 [OBS] \pm 0.005 [GAL] \right ) \times 10^{-12}$ and that
gives an agreement
$$
{\dot P_b ^{GR} \over \dot P_b ^{OBS}} = 1.0023 \pm 0.0041[OBS] \pm 0.021[GAL]
\eqno{(35)}$$
This is impressive and any deviation from general relativity due to any
other source of energy loss must be at most $\sim O(0.1 - 1)\%$ of the
general relativity predictions.

Most gravity theories alternative to general relativity, in particular
scalar-tensor theories, predict the existence of dipole gravitational
radiation as well \cite{wz,will}. In a binary system the magnitude of this
effect depends on the self-gravitational binding energies of the two
bodies. Following always ref.\cite{will}, the predicted energy loss rate can be
written 
$$
{dE \over dt} |_{D} = - {1\over3}{\kappa_D} < {\mu^2 m^2 \over r^4} {\cal E}^2>
\eqno{(36)}$$
where $\kappa_D$ depends on the theory in question and $\cal E$ is the
difference in the self-gravitational binding energies per unit mass
between the two bodies of the binary system. Expression (36) gives then
$$
{dE \over dt} |_{D} = - {1\over3}{\kappa_D}{\cal E}^2{\mu^2 m^2 \over \alpha^4}G(e) 
\eqno{(37)}$$
$$
{\dot P_b \over P_b} |_{D} = - {\kappa_D}{\cal E}^2{\mu m \over \alpha^3}G(e) 
\eqno{(38)}$$
where now
$$
G(e) = \left( 1 + {1\over2}e^2 \right) (1 - e^2)^{-5/2}
\eqno{(39)}$$

In the string theory under discussion, dipole terms come from the source
term $\Sigma$ in the field equation (13) of the massless dilaton.
Equivalently, it is the term  ${\cal L}_{int}  \sim  \alpha (\phi_0)  (\phi -
\phi_m)  {\bar {\psi_n}} \psi_n$ in the effective lagrangian for the
dilaton interaction with the nucleon field $\psi_n$ (as applied to the
macroscopic system of a binary pulsar) that leads to dipole radiation.
For our purposes it suffices to consider precisely this dipole radiation
effect. Either way, we find the expressions
$$
{dE \over dt} |_{D} \simeq - {4\alpha^2 (\phi_0) \over 3} \left( {\Omega_1
\over m_1} - {\Omega_2 \over m_2} \right)^2 {\mu^2 m^2 \over {\alpha^4}}G(e)
\eqno{(40)}$$
$$
{\dot P_b \over P_b} |_{D} \simeq - {4\alpha^2 (\phi_0) } \left(
{\Omega_1 \over m_1} - {\Omega_2 \over m_2} \right)^2{\mu m \over
\alpha^3}G(e)
\eqno{(41)}$$
where $\Omega_1$, $\Omega_2$ are the self-gravitational binding energies
of the two bodies. \footnote {Expressions (40)-(41) for the dipole radiation
are only approximate. Formula (41), for example, should be corrected by the
factors $\left [1 + \alpha_1(\phi) \alpha_2(\phi) \right]\left [1 +
{\beta_A \over {1 + \alpha^2 (\phi_0)}} \right ]^2$. Here, the first factor comes
from Kepler's third law, where the gravitational coupling is in fact an
effective gravitational constant between the two self-gravitating stars  with
$\alpha_1(\phi)$,$\alpha_2(\phi)$ strong-field modified values of
$\alpha(\phi)$ inside the pulsar and the its companion. The second factor is
due to the fact that the violation of the strong equivalence principle is
proportional to the PPN parameter $\eta \left (=4 \beta_{Edd} - \gamma_{Edd}
-3 \right)$. We thank G.Esposito-Farese for bringing this point to our
attention.} We now demand that the contributions (40)-(41) do not upset the
agreement (35) of general relativity with the observed data for the binary
pulsar PSR 1913 + 16. This means that these contributions must be within (0.1 -
1)$\%$ of the corresponding contributions (32)-(33)as applied in general
relativity. So, upon substituting the appropriate numerical values \cite{will},
we find for $\Delta \phi_{BP} \equiv  \left(\phi_0 - \phi_m \right)_{BP}$ the
bound $$ \Delta \phi _{BP} \lessim 10^{-\left(4-5\right)}  \eqno{(42)}$$ of the
same order of magnitude as the bound (29). So, both bounds from primordial
nucleosynthesis and the binary pulsar essentially coincide with each other and
with the attraction factor of the matter-dominated era as found in
ref.\cite{dp1}.  \bigskip

\noindent{\bf  V. Inflation}
\medskip
\nobreak

Since inflation \cite{guth} is a mechanism available in the early universe with
which large exponential enhancements or supressions are associated, it would be
interesting to see what is the effect of an inflationary epoch on the
relaxation shift of the dilaton. In fact, it is not difficult to see that
inflation could provide a very efficient early relaxation shift for the
dilaton of an order well below all the limits already discussed. Suppose that
the scalar field $\sigma$ in the action (6) plays the role of the inflaton  and
the potential $V(\sigma)$ satisfies the appropriate conditions for inflation.
Then, in the quadratic approximation (17) for $B(\phi)$ with $\Delta \phi =
\left(\phi - \phi_m \right)$, the field equation of $\phi$ gives ( upon
neglecting the spatial variations ) $$ {\Delta \ddot \phi} + 3 H {\Delta \dot
\phi} + {3\over2} H^2 \kappa \Delta \phi = 0
\eqno{(43)}$$
where $H^2 \simeq 2qV(\sigma)/3$ is the approximately constant Hubble
parameter during the slow-roll period of inflation. Equation (44) is
easily solved approximately by
$$
\Delta \phi \simeq e^{\left(-\frac{3}{2} \pm \sqrt{\frac{9}{4} -
\frac{3}{2}{\kappa}}\right)\int{H(t)}dt} \simeq e^{-c H \tau}
\eqno{(44)}$$
where c is some number $\sim O(1)$ and $\tau$ the period of inflation.
So, for an inflationary epoch satisfying the usual duration condition $H
\tau \grtsim 70$ we expect a quick relaxation shift of the order
$$
\Delta \phi_{INF} \lessim 10^{-30}
\eqno{(45)}$$
We see that, under some assumptions, inflation seems to be an extremely
effective mechanism driving the dilaton field $\phi$ very early and in a
quick and efficient way toward its equilibrium value $\phi_m$. If we then take
into account the total attraction factor found in ref.\cite{dp1}, we see that
today we expect an overall attraction factor ( due to inflation and the
subsequent radiation- and matter-dominated periods) of the order
$$
\Delta \phi_0 \simeq 10^{-40}
\eqno{(46)}$$
If this is true, it is evident that any deviations from general
relativity would be extremely suppressed beyond any potentially
observational limit.

A careful in depth analysis of inflation within the mechanism of ref.\cite{dp1}
has appeared very recently \cite{dv}. The
authors of ref.\cite{dv} also study the quantum creation of
dilatons during the primordial inflationary era and find that the
resulting quantum fluctuations are naturally compatible with
observational limits.
\bigskip

\noindent{\bf VI.\quad Discussion}
\medskip
\nobreak

The ``least coupling'' scheme introduced in ref.\cite{dp1} for driving the
dilaton vacuum expectation value toward values, which extremize the
strengths of its interactions, is a very appealing proposal. The initial
work of ref.\cite{dp1} has studied the cosmological evolution of the dilaton
field during the radiation- and matter-dominated eras of classical
cosmology and showed that it can be safely reconciled with cosmological
data. In the present work we have studied some additional cosmological
and astrophysical constraints for the shift of the dilaton toward its
equilibrium value in order to shed some light on how much close to that it
has been settled down by today. Apart from the big-bang nucleosynthesis
constraint (29), we have found a similar bound (42) for such a shift
derived from the emission of massless dilatons from the binary pulsar. We have
also seen that an inflationary era can easily drive
the dilaton very quickly to values safely below these limits and the overall
atraction factor of ref.\cite{dp1}. In fact, if inflation is operative, the
dilaton is driven to values which render the present scheme almost
indistinguishable from general relativity. An in depth study of inflation within
the mechanism of ref.\cite{dp1} has been done in ref.\cite{dv}.  \footnote {In
another development, general relativity has been also shown to be an attractor
of an arbitrary scalar-tensor theory in the context of stochastic inflation
\cite{gbw}.}

From the fundamental theory point of view, the present scheme relies on
some crucial assumptions. The basic assumption is that the various
dilaton coupling functions $B_i(\Phi)$ admit a local maximum at some
common value $\Phi_m$. This is guaranteed if all $B_i(\Phi)$ coincide
with a common function $B(\Phi)$, factorized as in (1) and admitting a
local maximum due to string loop effects as in (2). The crucial role
here is played by the symmetries of the underlying string theory. The
behaviour of the dilaton proper is related to the conjectured S-duality
symmetry of the string theory. On the other hand, the T-duality
symmetry, known to hold for some of the moduli fields, can be invoked
to guarantee that the same scheme is also applicable to them. The character
of these symmetries is lately a fruitful area of research and their relation
to the scheme under discussion remains to be seen. \footnote { It is
indicative to mention here the following. We have assumed that
inflation is driven by some field other than the dilaton ( or moduli in
general \cite{dv}; recent alternative schemes include dilaton
inflation \cite{vgv}, moduli inflation \cite{bbmss} and non-critical string
theory inflation \cite{emn}). Supersymmetry is generally broken by the
non-vanishing vacuum energy density present during inflation, thus lifting
the flat directions of the effective supergravity theory. Flat directions are
important for other reasons as well, e.g. for baryogenesis mechanisms. It is
then reassuring that a ``Heisenberg symmetry'' can be used in
the effective supergravity models to show that flat directions can be
preserved \cite{gmo}. The no-scale supergravity models, for which e.g. formula (11) is
applicable, are a special case of these models. This fact further underlines the
significance of some symmetries for the kind of theories we are
considering.}

Our results, more specifically the bound (42) from the binary pulsar,
were derived under the assumption that dilatons are massless. However,
once realized that inflation, if present, is an extremely efficient way
responsible for relaxing the dilaton toward its present value, one could
possibly start speculating about possible values for the masses of the
relevant scalar fields, namely the dilaton as well as the various
moduli. In fact, the authors of ref.\cite{dv} remark that a mass term is
possible and does not create the usual Polonyi \cite{cfkrr}- moduli \cite{dccqr,bbs}
problem ( associated with particles of mass $\sim O(M_W)$ and
gravitational strength couplings to ordinary matter, which dominate the
energy density of the universe until the temperature is too low for
nucleosynthesis to occur). This is simply achieved here because, after
inflation, the vacuum expectation value of the relevant field is left
very precisely at the place where it stores no potential
energy. Moreover, the same authors find that there is a very wide range
of masses exceeding $\sim 10 \rm GeV$ and extended up to the Planck
scale $M_P$, for which dilatons or moduli are essentially stable and
dominate the mass density of the universe, offering thus the possibility
of a novel type of dark matter.

However, there is no apparent reason to treat the dilaton on the same
footing as the other moduli. The dilaton is intimately connected with
the graviton and its vacuum expectation value determines the gauge
coupling constant at the string scale, whereas the moduli are associated
with compactified dimensions and their real part determines the radii of
them. There are cases with massless dilatons and massive moduli.
Moreover, some arguments have been put forward \cite{bbs,susskind} that a
mechanism
for cancelation of the cosmological constant requires a light weakly
scalar field - the dilaton - with a mass about the fourth root of the
observational bound of the cosmological constant-vacuum energy $\left(
10^{-46} \rm GeV^4 \right)^{1/4} \sim 10^{-3} \rm eV$. Depending on the
mechanism transmitting the supersymmetry breaking to the standard model,
there is the possibility of associating a light dilaton $m_\phi \sim {
m_{SUSY}}^2 / M_P$ of a mass of this order of magnitude ( corresponding to
a supersymmatry breaking scale $\sim (1-10) \rm  TeV$ ) with a
non-vanishing cosmological constant. In fact, there is a diverse set of
recent observations suggesting that the universe may possess a non-zero
cosmological constant \cite{kt}. We still need inflation to avoid
overclosure of the universe in the present epoch \cite{dv}. 
The possibility, however, of a light dilaton, associated with an interconnection
between a small vacuum energy of order e.g.
$exp(-c 4\pi/{g_{em}}^2) {M_P}^4$ ( compare with the structure of formulae
(9), (11) ) and a not completely exact symmetry responsible for
diminishing the cosmological constant and related to the present
scheme, remains as an intriguing one.

\newpage
\noindent {\bf Acknowledgements}
\medskip
\nobreak

I acknowledge informative discussions with C.Kounnas, a stimulating
talk by J.H.Taylor and the kind hospitality of CERN Theory Division. I am also
indebted to G.Esposito-Farese for valuable correspondence concerning the
interpretation of the binary pulsar data.

\end{document}